\documentclass{ws-ijmpd}
\usepackage[super,compress]{cite}
\usepackage{graphicx}

\begin{document}
\markboth{Arkhipova, Pilipenko}
{Dark energy tests at z\textgreater2}

%
\catchline{}{}{}{}{}
%

\title{Perspectives for studying the dark energy at z\textgreater2 with galaxy surveys}

\author{NATALIA A. ARKHIPOVA}

\address{Astro Space Center of P.N. Lebedev Physical Institute, Russia, 84/32 Profsoyuznaya str.,
Moscow, 117997 Russia\\
arna@asc.rssi.ru}

\author{SERGEY V. PILIPENKO}

\address{Astro Space Center of P.N. Lebedev Physical Institute, Russia, 84/32 Profsoyuznaya str.,
Moscow, 117997 Russia\\
spilipenko@asc.rssi.ru}

\maketitle

\begin{history}
\received{Day Month Year}
\revised{Day Month Year}
\end{history}





\begin{abstract}
{
Now creation of big catalogs of galaxies for measurement of baryon acoustic oscillation (BAO) is actively conducted. 
Existing and planned in the near future surveys are directed on the range of red shifts of z$\lesssim$2.
However, some popular models of dark energy (DE) give the maximum deviation from $\Lambda$CDM at z\textgreater2 
therefore we investigated sensitivity of hypothetical high redshift surveys to the model of DE. 
We have found that with the increase of the number density of detected galaxies at z\textgreater2 high 
redshift observations may give better constraints of DE parameters.
}
\end{abstract}
\ccode{PACS numbers:}

\section{Introduction}

Several cosmological observations show that
our universe is expanding with an acceleration\cite{Weinberg:2012es,Perlmutter:1999,Riess:1998,Riess:2007,Riess:2004,Albrecht:2006,Hicken:2009,Kessler:2009,Huterer:2010eh,Frieman:2008sn}.
This fact can be interpreted as a dominance of the energy of the unknown nature, so called {\it
dark energy} (DE)\cite{Linder:2010aa,Linder:2010ab,Lukash:2008,Wood-Vasey:2007,Wiltshire:2011ba}.
The main feature of this energy consists of
 negative pressure that leads to an accelerated
expansion. The standard cosmological scenario implies that order of
75\% of the total energy density is present in the form of DE. There
are several observational data based indications that DE is highly
spatial uniform and isotropic, as well as that the DE
became dominant recently. Definitely the nature of DE is one of
major puzzles of modern cosmology\cite{Linder:2010aa,Frieman:2008sn,Wiltshire:2011ba}.
A lot of theories of DE have been proposed\cite{Martin:2012bt,Linder:2008,Tsujikawa:2010sc}.

The simplest model of DE is the $\Lambda$CDM model, called a {\it
concordance model}, that assumes that the accelerated expansion of
the universe is driven by the presence of a cosmological
constant\cite{Weinberg:2008}. This model fits well the cosmological
observations, but the $\Lambda$CDM model has the coincidence and the
fine tuning still unexplained
problems\cite{Weinberg:2008,Peebles:2002gy}.

Instead of the considering the cosmological constant model there were several models proposed in which DE is a dynamical quantity and 
in these models DE is associated with a {\it dynamical scalar field} .
For the $\Lambda$CDM model the equation of state parameter $w=P_{\rm de}/\rho_{\rm de}$ ($P_{\rm de}$ is a pressure and $\rho_{\rm de}$ is an energy density of the DE) is a constant and it equals to minus one,
whilst  for the dynamical scalar field models  the equation of state parameter is a time varying function\cite{Peebles:2002gy}.

Depending on the value of the equation of state parameter at present, the time dependent DE models are divided into the phantom models\cite{Scherrer:2008be} ($w<-1$) and the {\it quintessence models}\cite{Caldwell:2005tm} ($-1<w<-1/3$).
   The quintessence models are subdivided into two classes: the thawing  models and the freezing (or tracking) ones.\cite{Chiba:2012cb,dePutter:2008,Zlatev:1999,Masiero:2000,Steinhardt:1999,Caldwell:1998,LaVacca:2009ee,Gupta:2011kw}

   In the tracking or freezing (slow roll) quintessence model the form of the potential allows the attractor  in the
late-time evolution of the scalar field be insensitive to the initial
conditions, and allows the scalar field energy density to track  the matter energy density in the matter domination epoch and then the radiation
energy density in the radiation domination epoch, remaining subdominant during these epochs. And only at late times, the scalar field becomes
dominant and starts behaving like a component with a negative pressure
driving the acceleration of the universe. Thus the quintessence models can clarify the coincidence problem.

   In this paper we have investigated the freezing quintessence model with an inverse power law Ratra-Peebles potential\cite{rp:88}:
 $V(\phi)=M^{\alpha+4}/\phi^{\alpha}$, $\alpha$ is a model parameter, defining the steepness of the potential; $\phi$ is a scalar field amplitude.

In order to distinguish between different dynamical DE 
models commonly constraint of energy equation of state $w(z)$ is
used, because different models of DE give different low of $w(z)$.
Recent Supernova Legacy Survey three year sample (SNLS3) combining
with other data on CMB, BAO and Hubble constant measurement gives
rise to $w_{de}=-1.013\pm0.068$ for constant of $w_{de}$ in standard
$\Lambda CDM$ models
\cite{Caldwell:2009ix,Suzuki:2011hu,Scovacricchi:2012jr,Wiltshire:2011ba}.

The BAO measurements the values of the equation of state parameter $w(z)$ ($z$ is a redshift) 
and its redshift derivative $dw/dz$ is the primary goal of the
ongoing  DE experiments such as SNLS3, VIPERS or BOSS, but
only the next generation of the large scale redshift surveys at $z\sim1$ and
beyond this  limit of the redshift like EUCLID\cite{Amendola:2012ys}, WFirst or BigBOSS\cite{March:2011rv} will be able
to provide the data to distinguish the DE models from each other.

We can get many information about the dynamical DE models analyzing the growth
of the matter perturbations which were
obtained from the redshift space distortion (RSD) surveys.
The classical quintessence models are not clustered, but they affect the
 rate of the matter evolution,  therefore the different
DE models predict the different growth rate history
\cite{Shi:2012ci,Gupta:2011kw,Samushia:2012,DiPorto:2011,Gong:2008,Pavlov:2012,Belloso:2011,Fu:2010,Lee:2010,Lee:2009,Samushia:2010ki}.
There are a lot of observational
growth rate data\cite{Di Porto:2007ym,Girones:2010,Hirano:2012,Xia:2009,Blake,Blake:2010}, but all these ongoing and future experiments are dedicated to the measurements in the range of the redshifts  $z<2$.

The main goal of our research is the estimation of
the sensitivity of the BAO and the RSD data to the cosmological parameters, especially to the values $w_{\rm de}(z)$ 
and the $dw_{\rm de}/dz$ in the range of the redshifts $z>2$.
Also we have explored what volume and number of the galaxies will be necessary to compete with the other surveys in the range of the redshifts $z>2$.

In this paper we will develop this ideas in quintessence model with Ratra-Peebles potential, that was well studied in many
papers\cite{avs:14ye, avs:15ye, pavlov:14ye,  pavlov:12ye}.

This paper is organized as follows:\\
The Introduction is presented in the Sec.~I.
In the Sec.~II we have considered a theory of the growth of matter perturbations for the Ratra-Peebles $\phi$CDM  model.
In the Sec.~III we have derived the responses of measured quantities to the DE model parameter $\alpha$.
In the Sec.~IV we evaluated the errors of BAO and RSD measurements.
Our discussions and conclusions are presented in the Sec.~V.

\section{Growth Factor of Matter Density Perturbations in Dark Energy Models}

The influence of the scalar field (of the Ratra-Peebles potential) on growth of structure
was well investigated in many papers\cite{avs:14ye, avs:15ye, pavlov:14ye, pavlov:12ye, taddei:15ye}.
Further we will follow the paper of O.~Avsajanishvili et. al.\cite{avs:14ye}
We use the linear perturbation equation for computation of the matter's overdensity\cite{Pace:2010, Campanelli:2011qd} $\delta$:
\begin{equation}
\delta^{''}+\Bigl(\frac{3}{a}+\frac{E^{'}}{E}\Bigr)\delta^{'}-\frac{3\Omega_{m,0}}{2a^{5}E^{2}}\delta=0
\label{deltaeq}
\end{equation}
where $\delta=\delta\rho_m/\rho_m$ is small perturbations in homogeneous universe expanding with the Hubble,
$\rho_m$ and $\delta\rho_m$ are the density and overdensity respectively, $\rho_{av}$ is average density of the universe.

A prime designates the differentiation with respect to a scale factor $a$,
$a=1/(1+z)$ where $z$ is a redshift; $E(a)=H(a)/H_0$ - the normalized value of the Hubble parameter $H(a)$ to a Hubble
constant $H_0 = 100h{\rm kms^{-1}Mpc^{-1}}$.

The first Fridmann equation in the case of the spatially-flat universe:
\begin{equation}
E^2(a) = \Omega_{\rm r,0}a^{-4} + \Omega_{\rm m,0} a^{-3} +
\Omega_{\rm \phi}, \label{ea}
\end{equation}
\noindent
where $E(a)$ --- total energy density of all matter components normalized to $H_0$,
$\Omega_{\rm r0}$ and $\Omega_{\rm m0}$ --- the radiation and the matter density parameters today. 
We have neglected the radiation term in the calculations because it is not relevant during the 
late stage expansion of the universe. We have applied  the fiducial values\cite{Ade:2013zuv}: $\Omega_{\rm m0}=0.315$, $h=0.673$.

The linear growth factor is defined as:
\begin{equation}
 D(a)=\frac{\delta(a)}{\delta(0)},
 \label{dz}
\end{equation}
  where $\delta(0)$ - a value of the density contrast at present.

The initial conditions are imposed as $D(a=1)=1$ and
$D(a=a_i)=a_i$, where $a_i=5\times10^{-5}$.

 The logarithmic derivative from the linear growth factor is called the growth rate:

\begin{equation}
G(a)=\frac{d\ln D(a)}{d\ln a}.
\label{f2}
\end{equation}

This function is parametrized as\cite{Wang:1998gt}:

\begin{equation}
 G(a) \approx \Omega^{\gamma}_{\rm m}(a),
 \label{f1f2}
\end{equation}
 where
\begin{equation}
\Omega_{\rm m}(a)=\frac{\Omega_{\rm m0}a^{-3}}{E^2}.
\label{fa}
\end{equation}

The $\gamma$ parameter is called a growth index. Suggesting the
correctness of general relativity,
 the value of $\gamma$ parameter can be evaluated  as\cite{Linder:2005in}:
 \begin{equation}
\gamma\approx0.55+0.05(1+w_0+0.5w_a),~~{\rm if}~~ w_0\ge-1
\label{gamma}
\end{equation}
where $w_0$ is a value of the equation of state parameter today and
$ w_a=(dw/dz)|_{z=0}=(-dw/ da)|_{a=1}$. For the $\Lambda$CDM model
$\gamma\approx0.55$.

\section{The Manifestation of the Quintessence Existence from the Observational Data}

In order to estimate the precision of the constraints of the value of the $\alpha$ parameter, we have found the response of the different cosmological tests on the change of the $\alpha$ parameter. For this purpose we have considered three kinds of cosmological tests which differ by their response on the change of the background dynamics and of the growth rate of the large scale structure:
\begin{enumerate}
\item
{\it the evolution of the Hubble parameter $H(z)$}.\\
 The values of the $H(z)$ can be obtained from the BAO (along the line of sight), SN Ia, the weak lensing data.
\item
{\it the luminosity distance and the angular size distance}. \\
Which values can be found from the BAO, SN Ia data.
 (In our analysis isn't relevant what kind of  distance is used and further we will use the common designation for distance $d(z)$).
\item
 {\it the linear growth factor $D(z)$ or the growth rate function $G(z)$}.\\ Which was obtained  from the RSD data (the bias factor isn't sensitive to the background changing).
\end{enumerate}

 The difference between the  measured values of the functions $H(z)$, $d(z)$, $G(z)$  and the corresponding  values of ones evaluated for the $\Lambda$CDM model may be by the evidence of the existence the quintessence.

 In order to illustrate this we have considered the quantity $\xi_X$ for the $\Lambda$CDM model:
\begin{equation}
\label{eq:xi} \xi_X = {1 \over X}\left. {\mathrm{d}X \over
\mathrm{d}\alpha}\right|_{\alpha=0}
\end{equation}
where $X$ may be one from the values $H$, or $d(z)$ or $G(z)$. When the value of the $\alpha$ parameter is small, the value of the $\xi_X$ is the relative difference of the quantity $X$ between the $\Lambda$CDM model and the Ratra-Peebles $\phi$CDM model. 

We also consider a quantity similar to  Eq.~\ref{eq:xi} for $\Omega_m$:
\begin{equation}
\label{eq:eta} \eta_X = {1 \over X}\left. {\mathrm{d}X \over
\mathrm{d}\Omega_m}\right|_{\Omega_m=\Omega_{m,0};\;\alpha=0}.
\end{equation}

\begin{figure}
\includegraphics[scale=1]{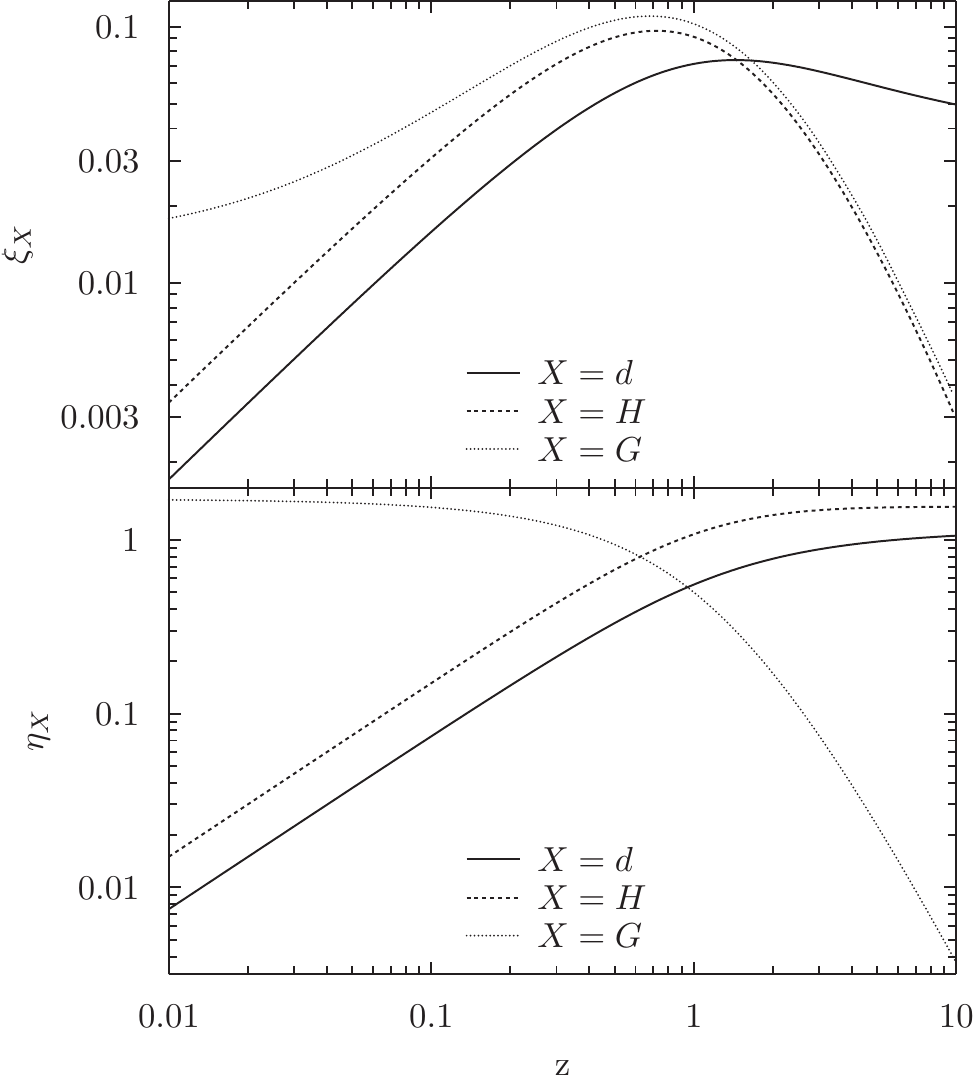}
\caption{The sensitivity of the measurements of $H(z)$, $d(z)$, $G(z)$ to the presence of the quintessence with a small $\alpha$ parameter (top panel) and to the change of $\Omega_{\rm m}$ parameter (bottom panel).}
\label{fig:alpha}
\end{figure}

As one can see from Fig.~\ref{fig:alpha}, the best response to $\alpha\ne 0$ can be reached for the redshift $z=0.7$ for $H(z)$ and $G(z)$ measurements and for the redshift $z=1.5$ for the distance $d(z)$ measurement. In the range of the redshifts $z>2$ the distance measurement of the $d(z)$ is the most sensitive.
However, these conclusions are based on the assumption that the present day values of the parameters $\Omega_{\rm m0}$ and $\Omega_{\rm de0}$ are evaluated perfectly well.  Since this is not the case, to verify the existence of the quintessence is necessary to investigate the differences between the measured values and the corresponding values of ones for the $\Lambda$CDM model.

In general, this problem should be solved by the MCMC simulations of observations for a set of different cosmological models, but for the better understanding it is instructive to analyze the two limiting cases. In the first case two quantities $X1$ and $X2$ ($H(z)$, or $d(z)$, or $G(z)$) are measured at a single redshift $z$. In the second case, the same quantity $X$ is measured at two different redshifts $z_1$ and $z_2$.

In order to analyze the first case, we have  considered a fiducial model with $\Omega_{\rm m}=\Omega_{\rm m0}$ and $\alpha=0$. We can apply the Taylor series expansion near the fiducial model for two measurements  $X1$ and $X2$:
\[
{\Delta X1 \over X1} = \eta_{X1}\Delta\Omega_m + \xi_{X1}\alpha,
\]
\[
{\Delta X2 \over X2} = \eta_{X2}\Delta\Omega_m + \xi_{X2}\alpha,
\]
eliminating $\Delta\Omega_{\rm m}$ and obtaining the value of the $\alpha$ parameter:
\[
\alpha = \left( {\Delta X1 \over \eta_{X1} X1} - {\Delta X2 \over \eta_{X2} X2} \right) \left/
\left( {\xi_{X1}\over \eta_{X1}} - {\xi_{X2}\over \eta_{X2}} \right) \right.
\]
Thus, when the quantities $X1$ and $X2$ are known with the fractional errors $\epsilon_{X1}$ and $\epsilon_{X2}$ respectively, the error for the calculation of the $\alpha$ parameter will be:
\begin{equation}
\label{eq:err2q}
\sigma_\alpha = \left( {\epsilon_{X1} \over \eta_{X1}} + {\epsilon_{X2} \over \eta_{X2}} \right) \left/
\left( {\xi_{X1}\over \eta_{X1}} - {\xi_{X2}\over \eta_{X2}} \right) \right.
\end{equation}

Similarly, when the same quantity $X$ is measured at two redshifts $z_1$ and $z_2$, we have obtained:
\begin{equation}
\label{eq:err2z}
\sigma_\alpha = \left( {\epsilon_X(z_1) \over \eta_{X}(z_1)} + {\epsilon_X(z_2) \over \eta_{X}(z_2)} \right) \left/
\left( {\xi_{X}(z_1)\over \eta_{X}(z_1)} - {\xi_{X}(z_2)\over \eta_{X}(z_2)} \right) \right. .
\end{equation}

Similar equations (\ref{eq:xi}-\ref{eq:err2z}) can be obtained from the well known Bayesian analysis with the Gaussian distributions and uniform priors or with the help of Fisher matrix, which is, in or terms:
\begin{align}
&F_{11} = {\xi_{X1}^2 \over\epsilon_{X1}^2} + {\xi_{X2}^2 \over\epsilon_{X2}^2},\\
&F_{12} = F_{21} = {\xi_{X1}\eta_{X1} \over\epsilon_{X1}^2} + {\xi_{X2}\eta_{X2} \over\epsilon_{X2}^2},\\
&F_{22} = {\eta_{X1}^2 \over\epsilon_{X1}^2} + {\eta_{X2}^2 \over\epsilon_{X2}^2}.
\end{align}

\section{Evaluation of the Relative Errors of the Cosmological Parameters Applying the BAO and the RSD Measurements.}
The question of the accuracy of the measurements of the cosmological parameters has been intensively studied in the recent few years\cite{Weinberg:2012es}.
In order to evaluate the relative errors of the BAO measurements  we have used the fitting formula\cite{Blake:2006}:
\begin{equation}
\epsilon=x_0\sqrt{{V_0 \over V}} \left( 1+ {n_0 (1+z)^2 \over n b^2} \right),
\label{eq:epsBAO}
\end{equation}
where $V_0=2.16 h^{-3}$ Gpc$^3$, $n_0=1.8\cdot 10^{-4}\, h^3$ Mpc$^{-3}$, $n$ --- number density of the galaxies in $h^3$ Mpc$^{-3}$, $b$ --- a galaxy bias factor at the scale of the BAO measurements, $x_0=0.85\%$ and $1.48\%$  for the distance $d(z)$ and the Hubble parameter $H(z)$, respectively.

\begin{figure}
\includegraphics[scale=1]{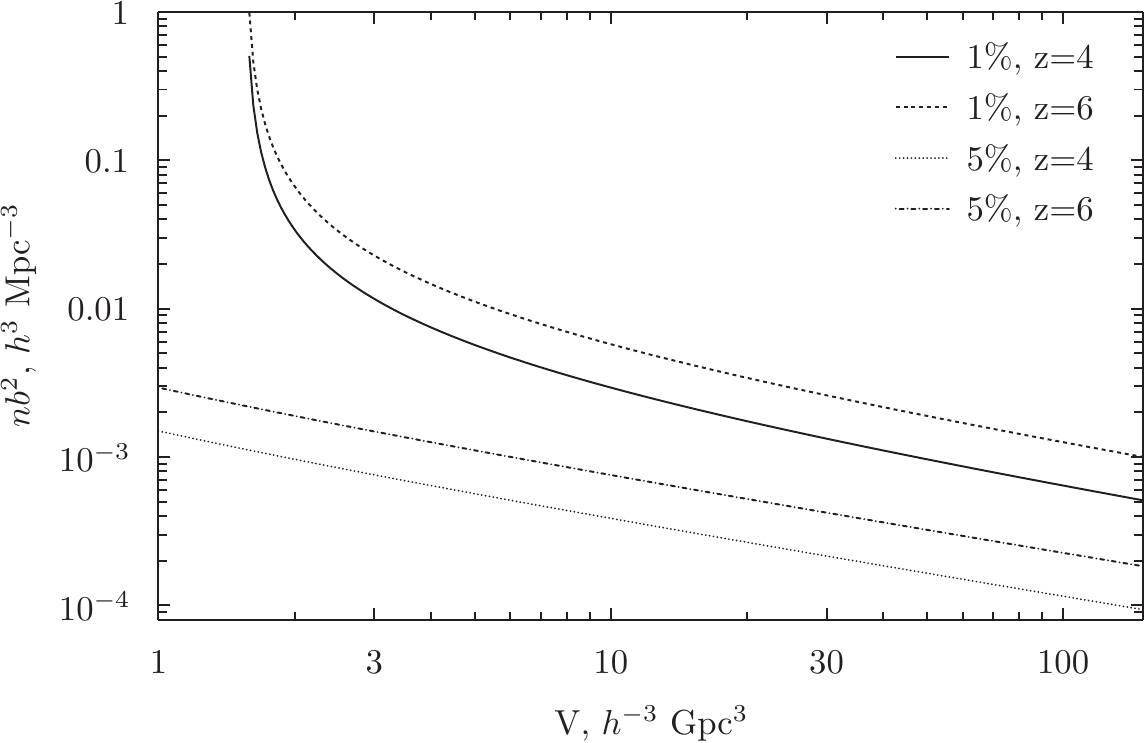}
\caption{The minimal comoving number density of galaxies $n$ as a function of survey volume $V$ needed to achieve precision of distance measurement $d(z)$ of 1\% or 5\% at $z=4$ and $z=6$ by the means of BAO observations.}
\label{fig:Vn}
\end{figure}

In order to reach the desired precision we have selected a pair of the quantities: $V$ and $n$. On Fig.~\ref{fig:Vn} we have shown the relation between these quantities for the precisions of the $d(z)$ measurements of 1\% and 5\% for the redshifts $z=4$ and $z=6$, respectively. For generality we have used  the combination $nb^2$. The bias depends on the kind of galaxies. For example, from the Planck mission results of measuring the Cosmic Infrared Background anisotropy\cite{PLANCK:2011}, for the submillimeter galaxies $b\approx 2.5$.

Concerning the RSD measurements, the precision can be estimated by the formula from Ref. \refcite{Guzzo:2008}:
\begin{equation}
\label{eq:epsRSD}
\epsilon_{RSD}={\delta \beta \over \beta}=\sqrt{V_0^{RSD} \over V} {1 \over n^{0.44}},
\end{equation}
where $V_0^{RSD}=0.025\,h^{-3}$ Gpc$^3$ and $\beta = G(z)/b$. The number density and volume
of a survey needed to reach the desired precision with the RSD measurements is
shown in Fig. 3.

\begin{figure}
\includegraphics[scale=1]{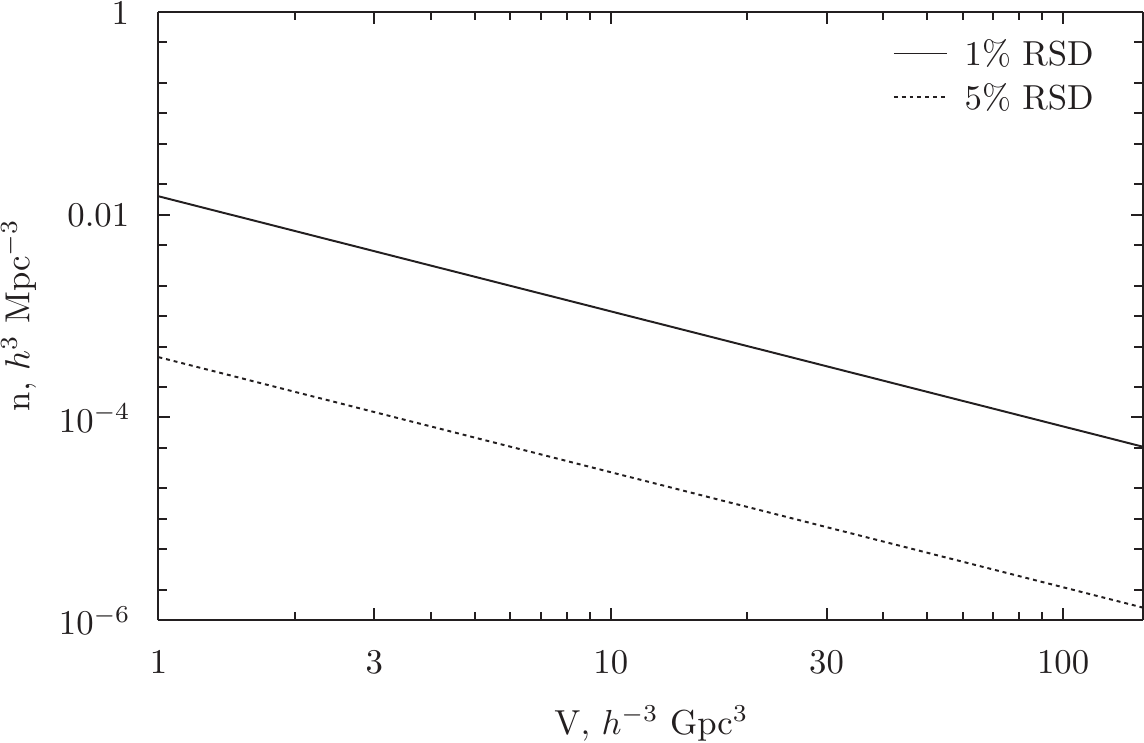}
\caption{The minimal comoving number density of galaxies $n$ as a function of survey volume $V$ needed to achieve precision of the growth factor $G(z)$ measurement of 1\% or 5\% by the means of RSD.}
\label{fig:VnRSD}
\end{figure}

We substitute the estimates  Eq.~\ref{eq:epsBAO} and Eq.~\ref{eq:epsRSD} for a hypothetical survey which covers all the sky and a logarithmic interval in the redshift in Eq.~\ref{eq:err2q} and  Eq.~\ref{eq:err2z} and analyze the results. we have assumed three choices of the number densities of the galaxies: $nb^2=10^{-2},\;10^{-3},\;10^{-4}$ for the BAO measurements and $n=10^{-3},\;10^{-3},\;10^{-4}$ for the RSD, in other words we have assumed $b\sim 3$.

\begin{figure}
\includegraphics[scale=1]{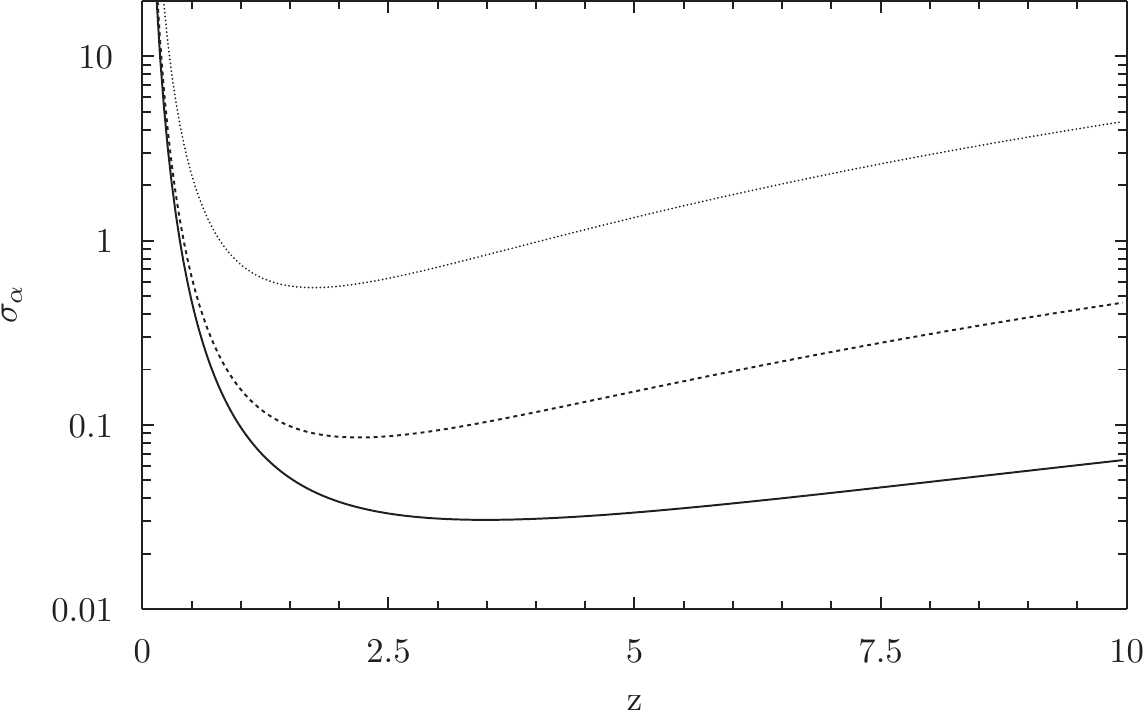}
\caption{The precision of the constraints on the $\alpha$ parameter from the BAO measurements of the $d(z)$ and the $H(z)$ functions at a single logarithmic redshift interval as a function of the redshift for number densities of the  galaxies $nb^2=10^{-2}$ (solid line), $10^{-3}$ (dashed line), $10^{-4}$ (dotted line).}
\label{fig:d_H}
\end{figure}

\begin{table}[ph]
\tbl{The existing and future spectroscopic galaxy samples for DE experiments}{
\begin{tabular}{lllll} \toprule
Survey & $N_\mathrm{gal}$ & Redshifts & Volume, $h^{-3}$Gpc$^3$ & $n$, $h^3$/Mpc$^3$ \\\colrule
2dFGRS\cite{colless:2001} & 221~414 & $z<0.3$ & 0.1  & $2\cdot10^{-3}$ \\
SDSS LRG\cite{percival:2007b} & 77~801 & $z<0.5$ & 1.5 & $5\cdot10^{-5}$ \\
BOSS (DR12)\cite{alam:2015} & 1~372~737 & $0.15<z<0.7$ & 15 & $9\cdot10^{-5}$ \\
WiggleZ\cite{drinkwater:2010} & $2\cdot10^5$ & $0.2<z<1$ & 3.6 & $5\cdot10^{-5}$ \\
BigBoss\cite{sholl:2012} & $2\cdot10^7$ & $z<1.7$ & 150 (14000 $deg^2$) & $1.3\cdot10^{-4}$ \\
DESI\cite{levi:2013} & $2\cdot10^7$ & $0.5<z<3.5$ & 430 (14000 $deg^2$) & $5\cdot10^{-5}$ \\
Wfirst & $10^7$ & $1.3<z<2.7$ & 50 (3400 $deg^2$) & $2\cdot10^{-4}$ \\
Euclid & $10^8$ & $0.7<z<2.0$ &  200 (15000 $deg^2$) & $5\cdot10^{-4}$ \\\botrule
\end{tabular}
\label{tab:cat}}
\end{table}

For the case of the measuring  pair of quantities at the same redshift, the resulting error of the $\alpha$ parameter is the smallest in the case of $d(z)- H(z)$ pair, and for the other two pairs the resulting error is higher on the order of the magnitude  for the redshifts $z\geq 1$ for the same galaxy number density. The values of the $\sigma_\alpha$ for different number densities of the  galaxies are shown on Fig.~\ref{fig:d_H}.

\begin{figure}
\includegraphics[scale=1]{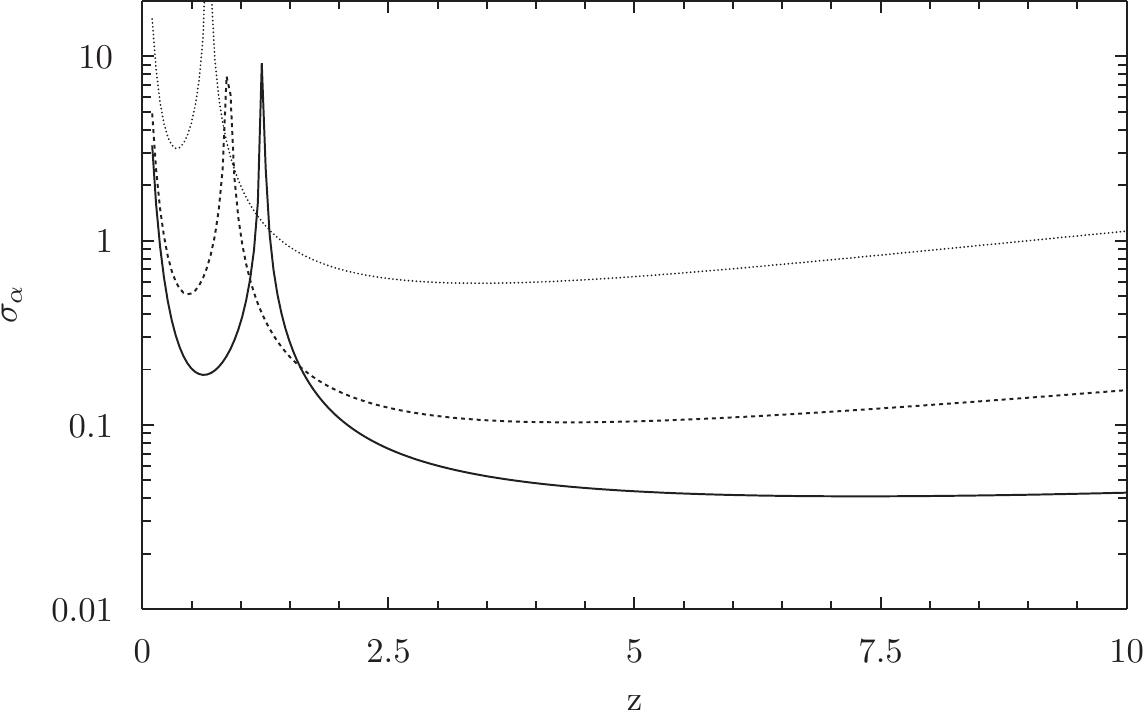}
\caption{The precision of the constraints on the $\alpha$ parameter from the BAO measurements of the $d(z)$ function in two logarithmic redshift intervals. The number densities of the galaxies $nb^2=10^{-2}$ (solid line), $10^{-3}$ (dashed line), $10^{-4}$ (dotted line) and the first redshift interval is centered at $z=1.22$, $z=0.87$ and $z=0.67$ respectively, while the position of the second one is  varied.}
\label{fig:d_z}
\end{figure}

\begin{figure}
\includegraphics[scale=1]{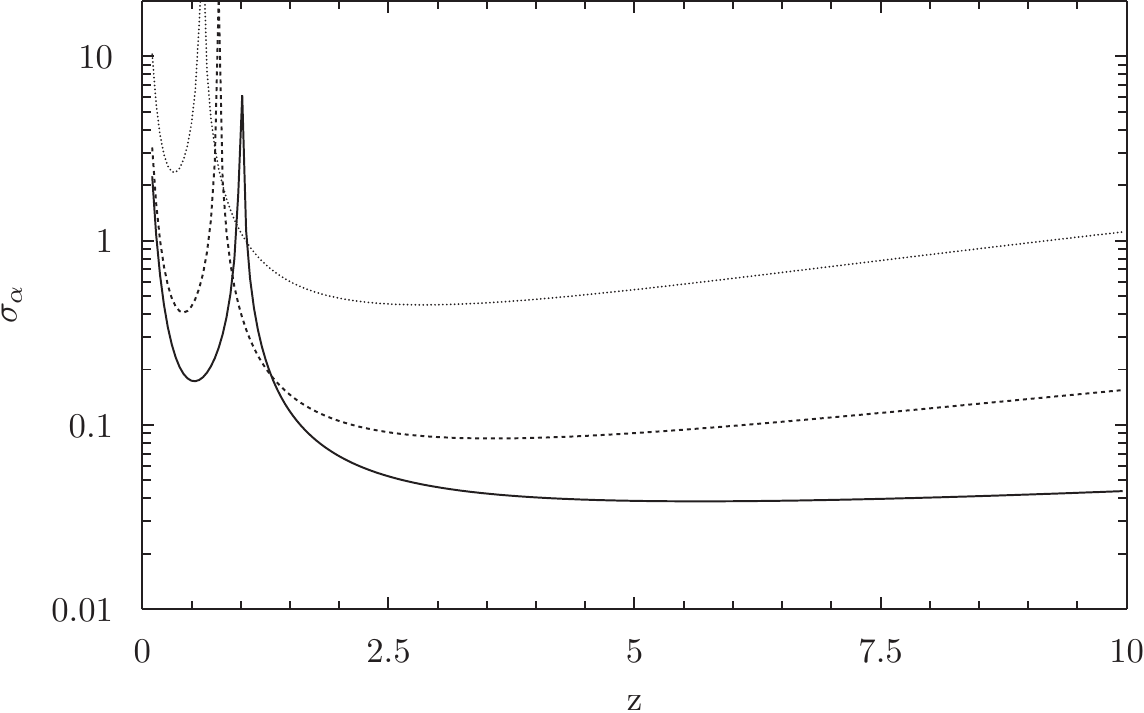}
\caption{The precision of the constraints on the $\alpha$ parameter from the BAO measurements of the $H(z)$ function  in two logarithmic redshift intervals. The number densities of the galaxies $nb^2=10^{-2}$ (solid line), $10^{-3}$ (dashed line), $10^{-4}$ (dotted line). The first redshift interval is centered at $z=1.0$, $z=0.77$ and $z=0.61$, respectively, while the position of the second one is varied.}
\label{fig:H_z}
\end{figure}

Secondly, the measurement of the of a single quantity at two different redshifts $z_1$, $z_2$ gives the most precise estimation when the $H(z)$ function or the $d(z)$ function is considered. For the $G(z)$ function the value of the $\sigma_\alpha$ is higher on the  order of the magnitude. We have varied both $z_1$ and $z_2$ and found the position of the global minimum of $\sigma_\alpha(z_1,z_2)$, $z_1<z_2$, and then we fix $z_1$ and plot $\sigma_\alpha(z1,z)$ on Figs.~ (\ref{fig:d_z}--\ref{fig:H_z}). Note that when $z_1=z_2$ there is insuffitient information to measure $\alpha$ and the error tends to infinity, thus, on Figs.~(\ref{fig:d_z}--\ref{fig:H_z}) the spikes are seen at $z=z_1$.

From the Table~\ref{tab:cat} one can see that the number densities  $nb^2\geq 10^{-3}$ will be available for the next generation catalogues constructed by the WFIRST and by the
EUCLID missions. For such number densities the parameters of these surveys are optimal for the constraining of the $\alpha$ parameter, (see the Figs.~(\ref{fig:d_H}--\ref{fig:H_z})). However, when the observations of a larger number of the galaxies at the high redshifts will become available and the density  $nb^2\geq 10^{-2}$ could be reached, the sensitivity of the constraints on the $\alpha$ parameter can be increased by  6-8 times by moving from the median redshift $z=1.5$ to the redshift $z\geq 3$.

The actual number density of galaxies in a survey depends on the luminosity function of galaxies and the detection limit of the survey. From the observed shape of the luminosity function\cite{bouwens:2015} at $z>4$ we conclude that in order to increase the number density of galaxies by one order of magnitude, the detection limit must be pushed down by 3--4 magnitudes.

Our estimates can be compared with the other predictions of the constraints on $\alpha$ by future missions, e.g. the prediction for Euclid\cite{amendola:2013} which gives $\sigma_\alpha = 0.055$ when the combined data of CMB and BAO measurements are used. We use equation (\ref{eq:err2z}) to estimate $\sigma_\alpha$. To take into account CMB measurements, we take the relative error in matter density $\epsilon_{\Omega_m} = 0.005$ and we place this measurement at $z=1100$. We also take into account that Euclid survey will cover roughly half of the sky. From our approach we have found $\sigma_\alpha = 0.07$ which is close to the result of Amendola et al.

\section{Discussion and Conclusions}
We have analyzed the observational manifestations of Ratra-Peebles model of dark energy at high redshifts, $z>2$, with the help of a simple approach similar to Fisher matrix. Our approach allows to find the precision of constraints on two model parameters, matter density $\Omega_m$ and spectral index of quintessence potential $\alpha$, from the accuracy of measurements of two observables. We have considered three observables: the Hubble parameter, the angular size distance and the growth rate of density perturbations. For these three observables we have considered two kinds of measurements: in the first case two different observables are measured at the same redshift. In the second case the same observable is measured at two different redshifts.

In order to make predictions on the accuracy of measurements of observables in future surveys we have taken from literature the approximations of the precision of BAO and RSD measurements.

We have considered all possible combinations of pairs of three observables listed above measured at a single redshift and also measurements of each observable at two different redshifts and found that at $z>2$ BAO measurements of are more sensitive to $\alpha$ than RSD.

We have found that when the number density of galaxies in future large surveys at these redshifts will approach to $nb^2\sim10^{-2}$, BAO measurements alone at $z>2$ will allow to improve constraints on parameter $\alpha$ in comparison with the measurements at $z\sim1.5$. This will require severe increase of the sensitivity of telescopes, so the missions aimed at measuring DE parameters from the large scale structure at $z>2$ need to be technically complicated than currently planned missions such as EUCLID.

\section*{Acknowledgments}
We thank for useful comments from Olga Avsajanishvili. This research was supported by Basic Research Program P-7 of the Presidium of the Russian Academy of Sciences and grant of the President of the Russian Federation for support of leading scientific schools of the
Russian Federation NSh-6595.2016.2. The work of Sergey Pilipenko is also partially supported by UNK FIAN.

\end{document}